\documentclass[aps,prl,amsmath,amssymb,twocolumn,showpacs,superscriptaddress,10pt]{revtex4-2}
\usepackage[english]{babel}
\usepackage{amsmath}
\usepackage{amssymb,amsfonts}
\usepackage{graphicx}
\usepackage[colorlinks=True,linkcolor=blue,citecolor=blue,urlcolor=blue]{hyperref}
\usepackage{bookmark}
\usepackage[dvipsnames]{xcolor}
\usepackage{braket}
\usepackage{nicefrac}
\usepackage{bm}
\usepackage{bbm}
\usepackage{slashed}
\usepackage[normalem]{ulem}
\usepackage{array}
\usepackage{makecell}
\newcolumntype{C}{>{$}c<{$}}
\usepackage{lipsum}
\usepackage{physics}
\usepackage{graphicx}
\usepackage{subfigure}
\usepackage{adjustbox}
\usepackage{bm}
\usepackage{bbold}
\usepackage[dvipsnames]{xcolor}
\usepackage{standalone}
\usepackage{multirow}
\usepackage{tikz}
\usepackage{dsfont}
\usepackage{float}
\usepackage{placeins}
\usepackage{comment}
\usepackage[utf8]{inputenc}
\usepackage[T1]{fontenc}
\usepackage{nicematrix}
\usepackage{leftindex}

\usepackage{siunitx}
\usepackage{physics}

\renewcommand{\Re}{\mathop{\mathrm{Re}}}
\renewcommand{\Im}{\mathop{\mathrm{Im}}}

\DeclareMathAlphabet{\zc}{OT1}{pzc}{m}{it}

  \DeclareUnicodeCharacter{2212}{-}

\usepackage{amsthm}

\newcommand{\ii}{\mathrm{i}}

\newcommand{\ipLR}[2]{\leftindex_\LL {\ip{#1}{#2}} _\RR}

\usepackage{booktabs}
\usepackage{multirow}

\newcommand{\LL}{\text{L}}
\newcommand{\RR}{\text{R}}

\usepackage{mathdots}

\begin{document}

\title{Non-Hermitian Landau Levels}

\author{Anton Montag}
\email{anton.montag@mpl.mpg.de}
\affiliation{Advanced Institute for Materials Research (WPI-AIMR), Tohoku University, Sendai 980-8577, Japan}
\affiliation{Max Planck Institute for the Science of Light, Staudtstraße 2, 91058 Erlangen, Germany}
\affiliation{Department of Physics, Friedrich-Alexander Universität Erlangen-Nürnberg, Staudtstraße 7, 91058 Erlangen, Germany}

\author{Tomoki Ozawa}
\email{tomoki.ozawa.d8@tohoku.ac.jp}
\affiliation{Advanced Institute for Materials Research (WPI-AIMR), Tohoku University, Sendai 980-8577, Japan}
\affiliation{RIKEN Center for Interdisciplinary Theoretical and Mathematical Sciences (iTHEMS), RIKEN, Wako, Saitama 351-0198, Japan}

\date{\today}

\begin{abstract}
    {We formulate non-Hermitian Landau levels in two-dimensional systems under a complex perpendicular magnetic field. In the symmetric gauge, we derive their discretely spaced, highly degenerate complex spectra and biorthogonal eigenstates, and clarify the role of non-unitary gauge transformations. A non-Hermitian Harper-Hofstadter lattice model confirms the continuum theory and reveals Gaussian wave packet dynamics governed by semiclassical equations with a complex Lorentz force, pointing to possible experimental realizations of complex magnetic fields.
    }
\end{abstract}

\maketitle

\emph{Introduction.}---Understanding the properties of a charged particle in a magnetic field has been pivotal for condensed matter physics.
In presence of a magnetic field in two dimension a charged particle exhibits equally spaced eigenenergies, called Landau levels~\cite{Landau2013}.
The presence of these highly degenerate Landau levels is responsible for various physical phenomena such as  de Haas–van Alphen effect~\cite{Ashcroft1976}, and the integer and fractional quantum Hall effects~\cite{Klitzing1980,Tsui1982}.
In addition, the discretized lattice description of magnetic fields in two dimension, the so-called Harper-Hofstadter model, is the paradigmatic model for Chern insulators~\cite{Thouless1982}.

The recent advances in non-Hermitian topology and the experimental development in non-Hermitian quantum mechanics has increased the interest in higher-dimensional non-Hermitian models~\cite{Bender2007,Rotter2009,Ashida2020,Bergholtz2021,Zou2021,Palacios2021,Shang2022,Yokomizo2022}.
The behavior and especially the topology of non-Hermitian systems have thereby been a focus of research, and the non-Hermitian analog of Chern insulators have been studied in detail~\cite{Kozii2017,Gong2018,Kawabata2019,Okuma2023}.
While there has been interest the response of non-Hermitian models to real-valued magnetic fields~\cite{Shen2018,Zhang2024,Kim2024,Teo2024,Wu2025,Longhi2025,Wang2026,Alon2026}, significantly less research has covered the effects of complex-valued magnetic fields~\cite{Paiva2023,Ozawa2024,Medina2025}.
The notable exception are purely imaginary vector potentials, which were introduced in the foundational Hatano-Nelson model and have been pivotal in describing the non-Hermitian skin effect~\cite{Hatano1996,Hatano1997,Yao2018,Brandenbourger2019,Borgine2020,Okuma2020,Ghatak2020,Helbig2020,Weidemann2020,Hofmann2020,Xiao2020,Xiao2021,Liang2022,Lin2023}.
The imaginary vector potential described the point-gap topology of one-dimensional systems~\cite{Kawabata2019,Bergholtz2021,Zhong2021} and the effects of imaginary vector potentials have been studied in two dimensions as well~\cite{Paiva2023,Ozawa2024}.
There the complex energy eigenvalues and the gauge independent Aharnov-Bohm damping have been discussed.

In spite of the importance of topology in the description of non-Hermitian systems and the examination of the change of conventional Landau levels upon the introduction of non-Hermiticity, the presence of non-Hermitian Landau levels in genuine complex-valued magnetic fields has largely been unexplored.
In this letter we show that equally spaced complex-valued eigenenergies arise if a two dimensional systems is subject to a complex-valued magnetic field. 
We derive both eigenenergies and the normalizable wavefunctions of the biorthogonal eigenstates of the continuum Hamiltonian under the symmetric gauge, labeled by level index and angular momentum. 
We show that, upon non-unitary gauge transformations to the Landau gauge, eigenstates may become not-normalizable depending on the values of the complex magnetic field.
We verify our results numerically and further show that the evolution of wave packet of the non-Hermitian Harper-Hofstadter model in the symmetric gauge is governed by the semi-classical equations of motion with complex-valued Lorentz force.

\emph{Non-Hermitian Landau levels.}---A charged particle confined to a plane with a perpendicular complex-valued magnetic field evolves under the Schrödinger equation
\begin{equation}
    \ii \hbar \partial_t \ket{\psi} = \hat{H} \ket{\psi} \, .
\end{equation}
The dynamics are governed by the non-Hermitian Hamiltonian
\begin{equation}\label{eq:Hamiltonian}
    \hat{H} = \frac{1}{2m} \left(\hat{\bm{p}} - \frac{q}{c} \bm{A}(\hat{x},\hat{y})\right)^2 \, ,
\end{equation}
with the canonical position and momentum operator denoted by $\hat{x},\hat{y}$ and $\hat{p}_{x,y}$, respectively, acting on all square integrable functions.
Here and throughout the letter vectors are indicated by bold symbols.
In the symmetric gauge the non-Hermitian vector potential given by $\bm{A}(\hat{x},\hat{y})=\frac{1}{2}B(\hat{x}\bm{e}_y-\hat{y}\bm{e}_x) $, where $\bm{e}_{x,y}$ are unit vectors in $x$ and $y$ direction.
We choose the coordinates such that $\Re(B)>0$.
As we point out below, our argument does not hold for the critical case of purely imaginary magnetic fields ($\Re(B) = 0$).
Introducing the complex-valued cyclotron frequency $\omega_c = \frac{qB}{mc}\in\mathbb{C}$ we define two distinct creation operators
\begin{subequations}\label{eq:creation}
    \begin{align}
        \hat{a}^\S &= \frac{1}{\sqrt{2m\hbar\omega_c}} \left[\frac{m\omega_c}{2}\left(\hat{x}-\ii\hat{y}\right) - \ii \left(\hat{p}_x-\ii\hat{p}_y\right)\right] \\
        \hat{b}^\S &=\frac{1}{\sqrt{2m\hbar\omega_c}} \left[\frac{m\omega_c}{2}\left(\hat{x}+\ii\hat{y}\right) - \ii \left(\hat{p}_x+\ii\hat{p}_y\right)\right]
    \end{align}
\end{subequations}
and the associated annihilation operators
\begin{subequations}\label{eq:annihilation}
    \begin{align}
        \hat{a} &= \frac{1}{\sqrt{2m\hbar\omega_c}} \left[\frac{m\omega_c}{2}\left(\hat{x}+\ii\hat{y}\right) + \ii \left(\hat{p}_x+\ii\hat{p}_y\right)\right] \\
        \hat{b} &= \frac{1}{\sqrt{2m\hbar\omega_c}} \left[\frac{m\omega_c}{2}\left(\hat{x}-\ii\hat{y}\right) + \ii \left(\hat{p}_x-\ii\hat{p}_y\right)\right].
    \end{align}
\end{subequations}
These operators fulfill the usual commutation relations
\begin{subequations}\label{eq:creationalgebra}
    \begin{align}
        &[\hat{a},\hat{a}^\S] = [\hat{b},\hat{b}^\S] = 1 \quad \text{and} \\
        &[\hat{a},\hat{b}^\S] = [\hat{b},\hat{a}^\S] =[\hat{a},\hat{b}] = [\hat{a}^\S,\hat{b}^\S] = 0 .
    \end{align}
\end{subequations}
We stress that $\hat{a}^\dagger\neq\hat{a}^\S$ and $\hat{b}^\dagger\neq\hat{b}^\S$, if the imaginary part of the magnetic field is nonzero.
Rewriting the Hamiltonian in terms of the creation and annihilation operators yields
\begin{equation}
    \hat{H} = \hbar\omega_c \left(\hat{a}^\S\hat{a}+\frac{1}{2}\right) \, , \label{eq:Hcreation}
\end{equation}
which is the complex-frequency analog of a quantum harmonic oscillator.
From the algebraic structure of Eqs.~\eqref{eq:creationalgebra} and \eqref{eq:Hcreation} that the eigenenergies of the non-Hermitian Landau level are given by
\begin{equation}
    \epsilon_n = \hbar\omega_c\left(n+\frac{1}{2}\right)
\end{equation}
and that each Landau level is infinitely degenerate, because the second pair of creation and annihilation operators do not appear in the Hamiltonian. 
As we discuss below, the eigenvectors corresponding to these eigenvalues are square-integrable.

\emph{Angular momentum.}---The two-dimensional angular momentum operator $\hat{L} = \hat{x}\hat{p}_y-\hat{y}\hat{p}_x$ commutes with the Hamiltonian in Eq.~\eqref{eq:Hamiltonian}.
Therefore they can be diagonalized simultaneously and we expect that the non-Hermitian Landau level eigenstates can be distinguished by considering the angular momentum.
Expressing the angular momentum operator in terms of the creation and annihilation operators results in
\begin{equation}
    \hat{L}=\hbar\left(\hat{a}^\S\hat{a}-\hat{b}^\S\hat{b}\right) \, .
\end{equation}
The angular momentum eigenvalues $l_{n,m} = \hbar\left(n-m\right)$ are determined by the algebraic structure of the creation and annihilation commutation relations Eq.~\eqref{eq:creationalgebra}.
Each eigenstate of the Hamiltonian is labeled the Landau level $n\in\mathbb{N}_0$, and an additional quantum number $m\in\mathbb{N}_0$.
\begin{figure*}
    \centering
    \includegraphics[width=\linewidth]{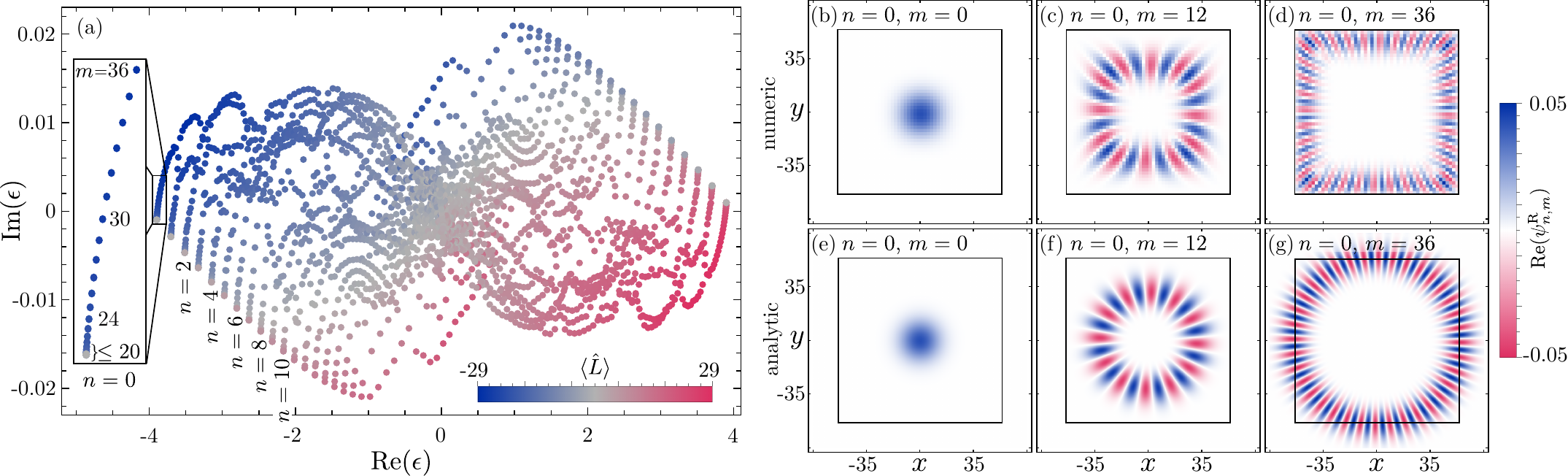}
    \caption{Eigenvalues and right eigenstates of the non-Hermitian Harper-Hofstadter model on a $50\times50$ square lattice with lattice spacing $a=1$ and the complex-valued magnetic field $B=0.1-0.001\ii$. In panel (a) the complex-valued eigenenergies are plotted, where the color indicates the planar angular momentum calculated from the eigenstates. The first ten non-Hermitian Landau levels can be identified from the eigenvalue distribution. The inset shows that for larger $m$ the numerical eigenenergy deviates from the analytical value as described in the text. To prove that this is indeed a finite-size effect we show the numerical eigenstates from three states from the lowest non-Hermitian Landau level in panels (b) to (d) and compare them to the wavefunctions given in Eq.~\eqref{eq:righteig}, which are plotted in panels (e) to (g). The black square is drawn in all plots along the edge of the simulated lattice. The comparison shows, that the Landau levels are recovered only as long as the wavefunctions are not extended beyond the edge of the simulated lattice.}
    \label{fig:spectrum}
\end{figure*}

\emph{Biorthogonal eigenstates.}---The eigenstates of the non-Hermitian Landau levels can be directly constructed from the creation and annihilation operators in Eqs.~\eqref{eq:creation} and \eqref{eq:annihilation}. 
We find the right eigenstates of the Hamiltonian 
\begin{equation}
    \ket{n,m}_\RR = \frac{\bigl(\hat{a}^\S\bigr)^n}{\sqrt{n!}} \frac{\bigl(\hat{b}^\S\bigr)^m}{\sqrt{m!}} \ket{0,0}_\RR \ ,
\end{equation}
where $\hat{H}\ket{n,m}_\RR = \omega_n \ket{n,m}_\RR$ and $\hat{L}\ket{n,m}_\RR = l_{n,m} \ket{n,m}_\RR$.
The right ground state $\ket{0,0}_\RR$ is annihilated by both $\hat{a}\ket{0,0}_\RR=0$ and $\hat{b}\ket{0,0}_\RR=0$.
The wavefunctions of the right eigenstates $\psi_{n,m}^\RR(x,y)=\ip{\bm{x}}{n,m}_\RR$ are given by 
\begin{widetext}
\begin{equation}\label{eq:righteig}
    \psi_{n,m}^\RR(x,y) = \frac{a_c^{n+m-1}}{\sqrt{2^{n+m+1}\pi n!m!}}e^{\frac{1}{4a_c^2}(x^2+y^2)}\left(\frac{\partial}{\partial x}+\ii\frac{\partial}{\partial y}\right)^m \left(\frac{\partial}{\partial x}-\ii\frac{\partial}{\partial y}\right)^n e^{-\frac{1}{2a_c^2}(x^2+y^2)}\, ,
\end{equation}
where we introduced the complex valued magnetic length $a_c=\sqrt{c\hbar/(qB)}$, with the branch cut of the complex square root taken along the negative real axis.
The wavefunctions are normalizable with respect to the norm $\iint dx\,dy |\psi_{n,m}^\RR(x,y)|^2$ because of the assumption $\mathrm{Re}(B) > 0$, and they are linearly independent.
However, due to the non-Hermiticity the right eigenstates are not orthogonal to each other.
Instead we need to construct the left eigenstates of the Hamiltonian, which are biorthogonal to the right eigenstates~\cite{Brody2013,Kunst2018}.
Left eigenstates are defined by $\leftindex_\LL{\bra{n,m}}\hat{H} = \omega_n \leftindex_\LL{\bra{n,m}}$ and in terms of the creation and annihilation operators they are given by
\begin{equation}
    \ket{n,m}_\LL = \frac{(\hat{a}^\dagger)^n}{\sqrt{n!}}\frac{(\hat{b}^\dagger)^m}{\sqrt{m!}} \ket{0,0}_\LL \, .
\end{equation}
The left eigenstates are derived from the left ground state, which is defined from the annihilation conditions $\bigl(\hat{a}^\S\bigr)^\dagger\ket{0,0}_\LL=0$ and $\bigl(\hat{b}^\S\bigr)^\dagger\ket{0,0}_\LL=0$.
The left wavefunctions $\psi_{n,m}^\LL(x,y)=\ip{\bm{x}}{n,m}_\LL$ are given by 
\begin{equation}\label{eq:lefteig}
    \psi_{n,m}^\LL(x,y) = \frac{\left(a_c^*\right)^{n+m-1}}{\sqrt{2^{n+m+1}\pi n!m!}}e^{\frac{1}{4\left(a_c^*\right)^2}(x^2+y^2)}\left(\frac{\partial}{\partial x}+\ii\frac{\partial}{\partial y}\right)^m \left(\frac{\partial}{\partial x}-\ii\frac{\partial}{\partial y}\right)^n e^{-\frac{1}{2\left(a_c^*\right)^2}(x^2+y^2)}\, .
\end{equation}
\end{widetext}
From Eqs.~\eqref{eq:righteig} and \eqref{eq:lefteig} it is clear that in the Hermitian limit $B\in\mathbb{R}$ the left and right wavefunctions are identical.
We have chosen the normalization of the ground state wavefunctions such that $\iint dx\,dy \, \left[\psi_{0,0}^\LL(x,y)\right]^*\psi_{0,0}^\RR(x,y)=1$.
The biorthogonality relation between the left and right eigenstates
\begin{equation}\label{eq:biorthogonal}
    \ipLR{n',m'}{n,m}=\delta_{n'n}\delta_{m'm}
\end{equation}
follows from the creation and annihilation operator commutation relations.

\emph{Non-unitary gauge transformations.}---In the derivation of the non-Hermitian Landau levels we chose the symmetric gauge as starting point of the calculation.
To obtain the Hamiltonian and its eigenstates in a different gauge we need to apply a gauge transformation leaving $B$ invariant.
However, the corresponding gauge transformations are not necessarily unitary for complex-valued magnetic fields.
Choosing for example the gauge function $\chi(x,y)=-(B/2)xy$ results in $\bm{A}(x,y)\rightarrow \bm{A}(x,y)-\nabla\chi(x,y)=Bx\bm{e}_y$,
which is commonly refereed to as Landau gauge~\cite{Landau2013b}.
The corresponding gauge transformation is given by $\hat{U}=\exp(-\ii \frac{q}{c\hbar}\chi(\hat{x},\hat{y})) = \exp(-\ii a_c^{-2}\hat{x}\hat{y})$~\cite{Landau2013}.
Since $a_c\in\mathbb{C}$ this constitutes a non-unitary gauge transformation $\hat{U}^{-1}\neq\hat{U}^\dagger$, which transforms the Hamiltonian of the system according to $\hat{H}\rightarrow \hat{U} \hat{H} \hat{U}^{-1}$.
Due to the unboundedness of $\hat{U}$ the wavefunctions of the transformed biorthogonal eigenstates $\ket{n,m}_\RR \rightarrow \hat{U}\ket{n,m}_\RR$ and $\ket{n,m}_\LL \rightarrow (\hat{U}^{-1})^\dagger\ket{n,m}_\LL$ may not be square integrable, and in such cases we can no longer consider $\epsilon_{n,m}$ as eigenvalues of the transformed Hamiltonian.
For the transformation to the Landau gauge it is straightforward to show that the wavefunctions $\Tilde{\psi}^\RR_{n,m}=\mel{\bm{x}}{\hat{U}}{n,m}_\RR$ scale as
\begin{equation}
    |\Tilde{\psi}^\RR_{n,m}|\sim\mathcal{O}([x,y]^{n+m})\, e^{-\frac{q}{4c\hbar}\left[\Re(B)(x^2+y^2)-2\Im(B)xy\right]} \, .
\end{equation}
Thus the wavefunctions are well localized and square integrable functions only for $\Re(B)>|\Im(B)|$.
If this condition is violated the wavefunctions are unbounded and the non-Hermitian Landau levels are not eigenvalues of the transformed Hamiltonian.
Even for $\Re(B)>|\Im(B)|$ the the distribution of the eigenstate wavefunction changes drastically between the symmetric and the Landau gauges.
Consequently, the choice of the gauge for complex-valued magnetic fields has physical consequences and must always be specified when discussing such models.

\emph{Non-Hermitian Harper-Hofstadter model.}---To confirm our description of the non-Hermitian Landau levels we numerically solve finite size non-Hermitian Harper-Hofstadter model with lattice spacing $a$~\cite{Harper1955,Hofstadter1976}.
For $a_c/a\gg1$ the flux through each plaquette of the lattice is small and we expect that the non-Hermitian Landau levels will be resolved.
The tight-binding Hamiltonian for the bare lattice without magnetic fields is given by
\begin{equation}
    \hat{H} = -\sum_{\langle i,j\rangle} t_{ij} \hat{c}_i^\dagger \hat{c}_j + \Tilde{t}_{ij} \hat{c}_j^\dagger \hat{c}_i 
\end{equation}
where $t_{ij}=\Tilde{t}_{ij}=1$ and the sum run over nearest neighbors.
For non-vanishing magnetic field each plaquette of the lattice is pierced by the complex-valued flux $\oint_{\scalebox{0.4}{$\square$}} \bm{A}(\bm{r})\cdot d\bm{r} = B$, where we set $a=1$.
We introduce the complex-valued flux by Peierls substitution
\begin{subequations}
    \begin{align}
        t_{ij} &\rightarrow t_{ij} \, e^{\ii\frac{q}{c\hbar}\int_{\bm{r}_j}^{\bm{r}_i}\bm{A}(\bm{r})\cdot d\bm{r}} \, ,\\
        \Tilde{t}_{ij} &\rightarrow \Tilde{t}_{ij} \, e^{-\ii\frac{q}{c\hbar}\int_{\bm{r}_j}^{\bm{r}_i}\bm{A}(\bm{r})\cdot d\bm{r}} \, .
    \end{align}
\end{subequations}
Transforming $t_{ij}$ and $\Tilde{t}_{ij}$ individually is important, because they are not related by complex conjugation for complex-valued magnetic fields.

We solve for the eigenvalues of the non-Hermitian Hofstadter model with $B=0.1-0.001\ii$ on a $50\times50$ lattice in the symmetric gauge with centered origin by means of exact diagonalization.
The complex-valued eigenenergies are plotted in Fig~\ref{fig:spectrum}.
Close to the lower boundary of the energy band, were the dispersion approximates the dispersion of the free continuum Hamiltonian from Eq.~\eqref{eq:Hamiltonian} we find the non-Hermitian Landau levels.
The eigenenergies cluster around $\epsilon=-4+\epsilon_n$.
To distinguish the different non-Hermitian Landau level eigenstates each eigenvalue is colored according to their angular momentum, calculated from the numerical right eigenstates.
While the non-Hermitian Landau levels are fully degenerate in energy for the continuum model, the imaginary part of the eigenvalues of the finite size Hofstadter Hamiltonian increases for decreasing angular momentum (larger $m$).
For each level this results in pronounced eigenvalue arcs that spiral into the center of the eigenvalue distribution.
This is an edge-effect of the finite-sized lattice model, which is more pronounced for lower angular momenta due to the increased spread of the wavefunction.
Along the edge the non-reciprocity of the hopping results in chiral edge amplification, similar to that of a periodic Hatano-Nelson model.
The Landau level eigenstates experience amplification for $\Im(B)<0$ and damping for $\Im(B)>0$, due to an interplay of angular momentum and the direction of the edge non-reciprocity.
In Fig.~\ref{fig:spectrum} three eigenstates are shown for the same non-Hermitian Landau level to confirm the relation between angular momentum and edge-induced gain of the eigenstate.
All three eigenstates are compared to the analytic solutions of the continuum model highlighting the differences between the wavefunctions at smaller angular momentum.

\emph{Wave packet dynamics.}---In addition to the diagonalization to the non-Hermitian Hofstadter model for weak complex-valued magnetic fields we simulated the evolution of a Gaussian wave packet on the lattice.

We prepared the wave packet initially centered at the origin of the lattice and imprinted a finite initial momentum $\bm{k}_0=\pi/40 \hat{\bm{e}}_x$ onto the wave packet. 
For a real magnetic field, the center of mass of the wave packet is described by semi-classical cyclotron motion, which forms a circle.
In contrast, the motion of the wave packet in the complex valued magnetic field is described by a spiral, which spirals inwards for $\Im(B)<0$ and outwards for $\Im(B)>0$.
The motion of the center of mass $\bm{r}_c$ is still described by the semi-classical equations of motion with complex-valued Lorentz force.
First, recall that, under the Hermitian magnetic field ($\mathrm{Im}(B) = 0$), the equation of motion for the center of mass position $\bm{r}_c = (x_c,y_c)$ under external electromagnetic fields $\mathbf{E}$ and $\mathbf{B}$ is $m_\text{eff}\ddot{\bm{r}}_c = q (\mathbf{E} + \dot{\bm{r}}_c \times \mathbf{B})$. 
In terms of the complex coordinate $z_c \equiv x_c + \ii y_c$, this equation can be written as $m_\text{eff}\ddot{z}_c = q (E_c - \ii B\dot{z}_c)$, where $E_c \equiv E_x + \ii E_y$. 
We find that the center-of-mass motion of a Gaussian wave packet under a complex magnetic field is described by the same equation where $B$ is replaced by a complex magnetic field. 
In terms of the original coordinates, this results in
\begin{equation}\label{eq:eom}
    m_\text{eff}\ddot{\bm{r}}_c = q\bigl[\bm{E}+\dot{\bm{r}}_c \times \Re(\bm{B}) + \Im(B)\dot{\bm{r}}_c\bigr] \, .
\end{equation}
From this expression, we can see that the imaginary part of the magnetic field results a drag force.
In the Supplemental Material~\cite{supmat}, we show how this equation can be recast as a complex frequency harmonic oscillator for vanishing electric field.
The frequency of oscillation is given by $\Re(\omega_c)$ and its imaginary part governs the spiraling behavior.
\begin{figure}
    \centering
    \includegraphics[width=\linewidth]{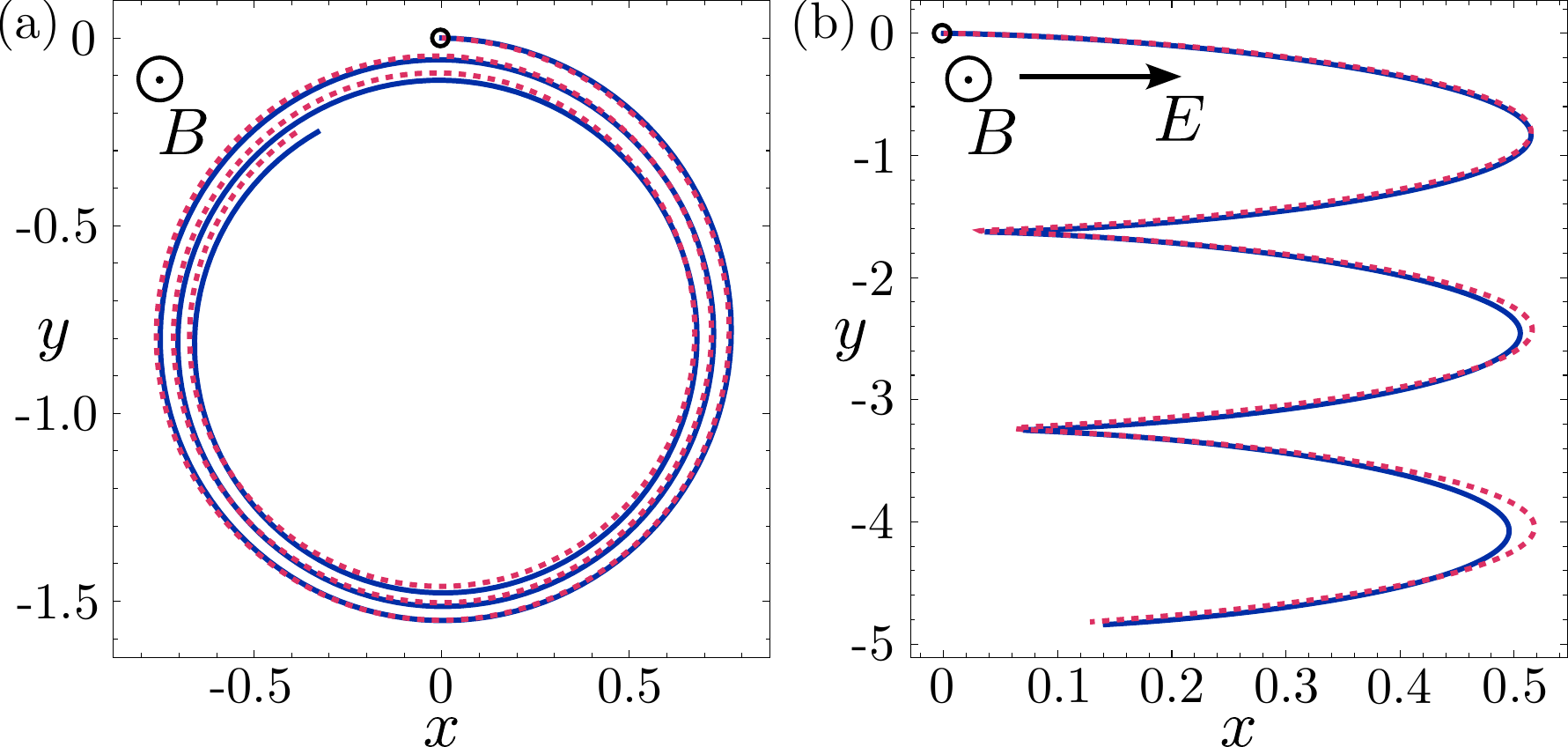}
    \caption{Trajectories of the center of mass of Gaussian wave packets in presence of a complex-valued magnetic field. Their evolution is simulated on a $50\times50$ square lattice governed by the Harper-Hofstadter model with $B=0.1-0.001\ii$ in the symmetric gauge with centered origin. The initial position is marked by a black circle. In panel (a) the spiral trajectory of a wave packet with finite initial velocity is shown in blue and compared to the solution of the semi-classical equation of motion, plotted in red. In panel (b) the wave packet is initially at rest and an in-plane electric field $\bm{E}=0.005\bm{e}_x$ is applied. The center of mass follows a skipping trajectory, shown in blue, and this curve is again compared to the analytic solution, shown in red. The analytic expressions are derived in the Supplemental Material~\cite{supmat}.}
    \label{fig:wavepacket}
\end{figure}
In Fig.~\ref{fig:wavepacket} the trajectory of the center of mass of the simulated wave packet is compared to the analytic solution, given the same initial conditions.
The deviations between the numerical solution and the analytical spiral can be explained by the width and position dependent effective mass $m_\text{eff}$ of the wave packet dynamics; see Supplemental Material for the details~\cite{supmat}.

Similar results are obtained for the wave packet initially at rest in a constant electric field $\bm{E}=E\hat{\bm{e}}_x$ with $E=0.1$ in addition to the complex-valued magnetic field.
The solution of Eq.~\eqref{eq:eom} is now given by a skipping trajectory, derived in the Supplemental Material~\cite{supmat}. 
This is confirmed by the numerical simulation of the dynamics, shown in Fig.~\ref{fig:wavepacket}.
Deviations from the analytic curve arise again due to the varying effective mass $m_\text{eff}$.

\emph{Experimental implementation.}---The non-Hermitian Landau levels can be measured in platforms that can implement non-Hermitian lattice models.
Prominent examples of such platform are topoelectric circuits  as well as acoustic, plasmonic and optical metasurfaces~\cite{Rafi2021,Ozdemir2019,Liu2022,Opala2023}. 
Then the spectral distribution of the full system or the evolution of wave packets on these lattices can be measured, and compared to our predictions.
Another possibility is to shape the desired Hamiltonian by employing adiabatic potentials~\cite{Berry1984,Dalibard2011,Lin2009,Lin2009b}. 
In systems with fast internal degrees of freedom, such as multilevel atoms in ultracold atomic gasses, the adiabatic evolution of said atoms is governed in part by adiabatic potentials, derived from the quantum geometry associated with the fast degrees of freedom.
If atoms in their excited state are lost from the condensate, these adiabatic potentials become non-Hermitian and thus can result in complex-valued magnetic fields acting on the condensate wave function~\cite{Montag2026}.

\emph{Discussion.}---Our work shows that complex-valued magnetic fields can be incorporated into continuum systems.
Considering the system in the symmetric gauge keeps essential properties such as the normalizability and locality of the eigenstate wavefunctions intact.
It further yields analytic expressions for the biorthogonal wavefunctions, which are labeled by Landau level index and angular momentum.
While we find a particularly well-behaved description of non-Hermitian Landau levels in the symmetric gauge, spectral and dynamical properties of other gauges are yet to be explored. 
We would also like to point out that, although it is known that under the real magnetic fields, the discrete Landau level eigenvalues exhaust the spectrum~\cite{AvronHerbstSimon1978}, we do not have a mathematical argument if the discrete complex Landau level eigenvalues we found exhaust the spectrum or if there should be additionally a continuous spectrum. 
The non-Hermitian Landau level wavefunctions allows us to construct certain many-body wavefunctions such as non-Hermitian Laughlin states, which potentially results in fractional topological states in non-Hermitian interacting systems.

\acknowledgments

A.M. would like to thank Flore K. Kunst for providing part of the funding and enabling his research stay at AIMR. A.M. acknowledges funding from both the Max Planck Society Lise Meitner Excellence Program~\mbox{2.0} and the FY2025 JSPS Postdoctoral Fellowships for Research in Japan (Short-term(PE)) with the ID No. PE25273. 
T.O. acknowledges financial support from JSPS KAKENHI Grant No. JP24K00548, JST PRESTO Grant No. JPMJPR2353, and JST PRESTO Convergence Research Grant No. JPMJCR26XA.

\bibliography{references.bib}

\clearpage

\begin{widetext}
\begin{center}
    {\large \textbf{Supplemental Material: Non-Hermitian Landau Levels}}
\end{center}

\section{SM1: Special solutions of the semiclassical equations of motion for the complex-valued Lorentz force}

Here we present the solutions of the semiclassical equations of motion for the wave packet center of mass dynamics presented in the letter for both numerically probed configurations.
The semi-classical equations of motion in presence of complex-valued magnetic fields, taken from Eq.~(17) in the letter, are given by 
\begin{equation}\label{eq:eom}
    m_\text{eff}\ddot{\bm{r}}_c = q\bigl[\bm{E}+\dot{\bm{r}}_c \times \Re(\bm{B}) + \Im(B)\dot{\bm{r}}_c\bigr] \, .
\end{equation}
The center of mass position of the wave packet is given by $\bm{r}_c=(x_c,y_c)^T$.
In the following we consider first the solution for vanishing electric field and include electric fields later.

\subsection{Spiraling cyclotron orbit}
For vanishing electric field and real magnetic field one obtains circular motion with the frequency $\omega_c$. 
If the magnetic field has finite imaginary part, we can show that this still holds, now with a complex frequency.
Rewriting Eq.~\eqref{eq:eom} by introducing a single complex coordinate $z_c=x_c+\ii y_c$ we obtain the differential equation
\begin{equation}\label{eq:complex_velo}
    \ddot{z}_c = -\ii \frac{qB}{m_\text{eff}} \dot{z}_c \implies \dot{z}_c= v_{z,0} e^{-i\omega_ct} \, ,
\end{equation}
where we set $c=1$ and $v_{z,0}=\dot{x}_c(0)+\ii\dot{y}_c(0)$ is a constant depending on the initial velocity of the wave packet.
We find the center of mass position by simply integrating Eq.~\eqref{eq:complex_velo} resulting in
\begin{equation}\label{eq:spiral}
    z_c= \ii\frac{v_{z,0}}{\omega_c} e^{-i\omega_ct} +\left(z_c(0)-\ii\frac{v_{z,0}}{\omega_c}\right)=\ii\frac{v_{z,0}}{\omega_c} e^{-i\Re(\omega_c)t}e^{\Im(\omega_c)t} +\left(z_c(0)-\ii\frac{v_{z,0}}{\omega_c}\right)\, .
\end{equation}
Since the frequency of the oscillation $\omega_c\in\mathbb{C}$ the trajectory always spirals inwards or outwards depending on $\Im(B)$, which can be seen by splitting the frequency in real and imaginary part.
We recover the center of mass position by $\Re(z_c)=x_c$ and $\Im(z_c)=y_c$.

To compare the analytic solution Eq.~\eqref{eq:spiral} to the numerical solution we numerically determined initial velocity and measured the effective mass of the bare lattice given the initial width of the wave packet. 
The initial position is given from the initiation of the wave packet, however due to the complex-valued vector potential there is no simple relation between the momentum imprinted on the wave packet and the initial velocity, thus $v_{z,0}$ needed to be determined numerically.

\subsection{Skipping orbit in electric field}
To solve Eq.~\eqref{eq:eom} in presence of a constant electric field we again introduce the complex coordinate $z_c=x_c+\ii y_c$ resulting in
\begin{equation}\label{eq:complex_velo2}
    \ddot{z}_c = -\ii \frac{qB}{m_\text{eff}} \dot{z}_c + \frac{qE_c}{m_\text{eff}} \, ,
\end{equation}
where $E_c=E_x+\ii E_y$.
We can solve for $\dot{z}_c$ by variation of parameters finding
\begin{equation}
    \dot{z}_c = -\ii \frac{E_c}{B} + \left(v_{z,0}+\ii \frac{E_c}{B}\right) e^{-i\omega_ct}
\end{equation}
Integrating this over $t$ and considering the initial conditions results in
\begin{equation}
    z_c = -\ii \frac{E_c}{B}t + \ii\frac{1}{\omega_c}\left(v_{z,0}+\ii \frac{E_c}{B}\right) e^{-i\omega_ct} + \left[z_c(0) -\ii\frac{1}{\omega_c}\left(v_{z,0}+\ii \frac{E_c}{B}\right)\right] \, ,
\end{equation}
where the first term described the drift perpendicular to the electric field and the second term results in the skipping motion with decreasing amplitude.
We again recover the center of mass position by $\Re(z_c)=x_c$ and $\Im(z_c)=y_c$.

In the numerical simulation the wave packet is initially at rest, therefore we do not need to fit the initial velocity but simply can set it to $v_{z,0}=0$.
The result is shown in Fig.~2 of the letter.

\section{SM2: Discussion of width dependent effective mass of wave packet}
As mentioned in the letter, the effective mass of the wave packet $m_\text{eff}$ depends in general on the width of the wave packet $\sigma$ and for finite $\Im(B)$ it is also (weakly) position dependent.
In the parameter regime used for the simulation the width of the wave packet effects the effective mass stronger. 
To highlight this we measured the effective mass of Gaussian wave packets on a $50\times50$ square lattice without any magnetic fields.
We prepared Gaussian wave packets at the center of the lattice with vanishing initial velocity and applied a constant electric field $\bm{E}=E\bm{e}_x$, accelerating the wave packet. 
From the numerically determined acceleration rate $a(\sigma)$ we can determine the effective mass as $m_\text{eff}(\sigma)=E/a(\sigma)$.
\begin{figure}
    \centering
    \includegraphics[width=0.5\linewidth]{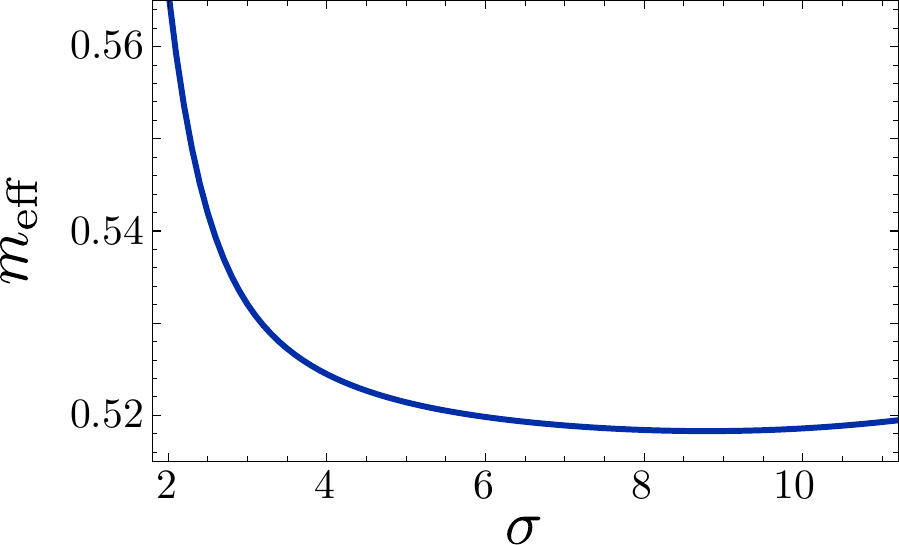}
    \caption{Effective mass $m_\text{eff}$ depending on wave packet width for a $50\times50$ square lattice for vanishing magnetic field.}
    \label{fig:mass}
\end{figure}
We show the results in Fig.~\ref{fig:mass}.

During the evolution in magnetic fields the width of wave packets changes periodically, which is sometimes called breathing of the wave packet~\cite{Tsuru1992,Ozawa2017}.
For real magnetic fields this change in width results in a varying velocity of the wave packet along the cyclotron orbit, but the orbit itself is not changed.
Due to the damping induced by the imaginary part of the complex-valued cyclotron frequency this is not the case for complex-valued magnetic fields. 
If the damping rate is modulated the trajectory changes, resulting in deviations between the numerical and analytical trajectories.

In addition to the width dependent mass changes there is a small position dependency of the effective mass, given finite $\Im(B)$.
However this position dependence is, in the position range explored by the wave packet in our simulation, about a magnitude smaller than the width dependence.
We thus believe that the oscillating width of the wave packet is responsible for the deviations of the numerically determined trajectory from the analytic expressions derived above.
\end{widetext}

\end{document}